\documentclass[pra,aps,twocolumn,notitlepage,superscriptaddress]{revtex4-2}

\usepackage[
pdffitwindow=true,
colorlinks=true,
frenchlinks=false,
linkcolor=blue,
anchorcolor=blue,
citecolor=blue,
filecolor=blue,
urlcolor=blue,
bookmarks=true,
bookmarksopen=true,
bookmarksnumbered=true,
bookmarksopenlevel=1,
plainpages=false,
pdfpagelayout=TwoPageLeft,
pdfpagelabels=true,
breaklinks
]{hyperref}
\usepackage[per-mode=symbol,separate-uncertainty]{siunitx}
\usepackage{graphicx}
\usepackage{dcolumn}
\usepackage{amssymb}
\usepackage{url}
\usepackage{color}
\usepackage{graphicx,wrapfig,lipsum}
\usepackage{bm}
\usepackage[english]{babel}
\usepackage{color}
\usepackage{ulem}
\usepackage{lipsum}
\bibpunct{[}{]}{,}{n}{}{}
\usepackage{braket}

\usepackage{chemformula}

\begin{document}

\title{Probing an electron spin ensemble with squeezed microwave signals\\ }
\author{P.\,Oehrl}
\affiliation{Walther-Mei{\ss}ner-Institut, Bayerische Akademie der Wissenschaften, 85748 Garching, Germany} 
\affiliation{Technical University of Munich, TUM School of Natural Sciences, Physics Department, 85748 Garching, Germany}

\author{F.\,Fesquet}
\affiliation{Walther-Mei{\ss}ner-Institut, Bayerische Akademie der Wissenschaften, 85748 Garching, Germany}
\affiliation{Technical University of Munich, TUM School of Natural Sciences, Physics Department, 85748 Garching, Germany}

\author{K.\,E.\,Honasoge}
\affiliation{Walther-Mei{\ss}ner-Institut, Bayerische Akademie der Wissenschaften, 85748 Garching, Germany}
\affiliation{Technical University of Munich, TUM School of Natural Sciences, Physics Department, 85748 Garching, Germany}

\author{M.\,Handschuh}
\affiliation{Walther-Mei{\ss}ner-Institut, Bayerische Akademie der Wissenschaften, 85748 Garching, Germany}
\affiliation{Technical University of Munich, TUM School of Natural Sciences, Physics Department, 85748 Garching, Germany}

\author{A.\,Marx}
\affiliation{Walther-Mei{\ss}ner-Institut, Bayerische Akademie der Wissenschaften, 85748 Garching, Germany}

\author{R.\,Gross}
\affiliation{Walther-Mei{\ss}ner-Institut, Bayerische Akademie der Wissenschaften, 85748 Garching, Germany}
\affiliation{Technical University of Munich, TUM School of Natural Sciences, Physics Department, 85748 Garching, Germany}
\affiliation{Munich Center for Quantum Science and Technology (MCQST), 80799 Munich, Germany}

\author{K.\,G.\,Fedorov}
\affiliation{Walther-Mei{\ss}ner-Institut, Bayerische Akademie der Wissenschaften, 85748 Garching, Germany}
\affiliation{Technical University of Munich, TUM School of Natural Sciences, Physics Department, 85748 Garching, Germany}
\affiliation{Munich Center for Quantum Science and Technology (MCQST), 80799 Munich, Germany}

\author{H.\,Huebl}
\email[]{hans.huebl@wmi.badw.de}
\affiliation{Walther-Mei{\ss}ner-Institut, Bayerische Akademie der Wissenschaften, 85748 Garching, Germany}
\affiliation{Technical University of Munich, TUM School of Natural Sciences, Physics Department, 85748 Garching, Germany}
\affiliation{Munich Center for Quantum Science and Technology (MCQST), 80799 Munich, Germany}

\date{\today}
\pacs{}
\keywords{} %Before starting of \begin{abstract}

\begin{abstract}
The efficient transfer of quantum states into a long-lived storage unit such as  solid-state spin ensembles is widely recognized as a critical challenge with significant implications for quantum communication, sensing and computing applications. Here, we experimentally investigate the interaction of propagating squeezed microwaves with an electron spin resonance transition in order to evaluate the use of spin ensembles as quantum memories for GHz signals. We generate continuous variable microwave states  with a squeezing  of up to $5~$dB below the vacuum level and let this signal interrogate a spin ensemble, which is inductively coupled to a lumped element superconducting microwave resonator with a cooperativity of $C=0.3$. Analyzing this signal using Wigner tomography, we observe a transfer efficiency of around 61\,$\%$ between the squeezed microwaves and the spin excitation. We successfully model our experimental results with a dedicated steady-state model based on the quantum input-output formalism and provide guidance for design parameters required to enable spin-based quantum memories.
\end{abstract}

\maketitle

%Introduction 
\textbf{Introduction.} The realization of secure communication in future quantum networks requires a high-fidelity transfer and storage of quantum states~\cite{gisin2007quantum, kimble2008quantum, bhaskar2020experimental, teller2025solid}. In particular, long-range quantum key distribution and quantum repeater schemes are expected to rely on long-lived quantum memories (QMs) to preserve nonclassical or entangled states and offset inevitable temporal delays in practical networks~\cite{zhang2022device, nadlinger2022experimental, pu2021experimental}. Another promising application of QMs in communication protocols concerns the idea of quantum tokens, where quantum states are associated with unique identities or credentials, which can be securely verified through communication with a trusted node~\cite{wiesner1983conjugate, pastawski2012unforgeable, jirakova2019experimentally}. Security of such schemes relies on fundamental information-theoretic assumptions, which provide access to the concept of unconditional security, in contrast to the computational security provided by modern classical encryption schemes.

Among various QM platforms~\cite{specht2011single, tang2015storage, wallucks2020quantum}, solid state spin ensembles are considered to be very promising candidates due to their long coherence times and ability to be coupled to both optical and microwave signals. In the last decade, a significant progress towards the implementation of long-lived QMs with spin ensembles has been achieved by demonstrating long coherence times up to several hours~\cite{bar2013solid, zhong2015optically}, the potential for scalability and for integration into networks~\cite{ghirri2016coherently, strinic2025broadband}, the ability to store and retrieve coherent states in the limit of single photons~\cite{timoney2013single, grezes2015storage}, and the realization of a random-access QM with classical microwave signals~\cite{o2022random}. Nonetheless, the realization of efficient quantum state storage and retrieval at microwave frequencies in the single photon regime remains an open challenge~\cite{eichler2017electron, bienfait2017magnetic, o2022random}. The experimental demonstration of such a microwave QM platform would enable further integration with superconducting quantum circuits, one of the leading architectures in quantum information processing. 

Here, we present experimental results based on a modular combination of a Josephson parametric amplifier (JPA) for the generation of squeezed microwave signals and a spin ensemble coupled to a microwave cavity for the QM implementation. We demonstrate the successful generation of propagating microwave squeezed states with squeezing levels up to $5.3$\,dB below vacuum. We use these nonclassical photonic states to couple them to the excitations of a spin-resonator hybrid and study the resulting response in the steady-state regime. This allows us to estimate a corresponding transfer efficiency of $61\%$ for the transduction of quantum microwave states into spin excitations. \\

\begin{figure}[t!]
    \centering
    \includegraphics[width=1\linewidth]{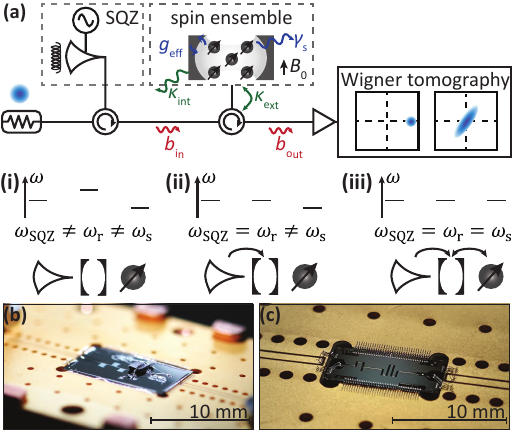}
    \caption{\textbf{Experimental setup.} (a) By sending a vacuum state to the phase-sensitively operated JPA, squeezer (SQZ), we generate a propagating squeezed vacuum signal, $\hat{b}_\mathrm{in}$ with a frequency $\omega_{\text{SQZ}}$, which is directed to the spin-resonator system via superconducting coaxial cables and a cryogenic circulator. The transfer of the squeezed signal to the spin excitation is mediated via a superconducting lumped-element microwave resonator. The output signal from the spin-cavity system, $\hat{b}_\mathrm{out}$, propagates further through a cascade of cryogenic and room temperature amplifiers, is then down-converted and digitized to perform the Wigner tomography of microwave signals. By tuning the static magnetic field $B_0$, we can distinguish three coupling cases: (i) the SQZ is off-resonant from the spin-resonator hybrid; (ii) the SQZ is in resonance with the resonator, but off-resonant with the spin ensemble; (iii) the SQZ, resonator and spin ensemble are tuned in resonance. (b) Photo of the spin-resonator system. (c) Photo of the JPA device.}
    \label{fig1:concept setup}
\end{figure}

\textbf{Experimental setup.}
A conceptual scheme of our experimental setup is presented in Fig.\,\ref{fig1:concept setup}. We generate squeezed vacuum states at microwave frequencies using a reflection-type, flux-driven JPA consisting of a coplanar waveguide resonator coupled to a nonlinear Josephson junction circuit~\cite{zhong2013squeezing, pogorzalek2017hysteretic, renger2021beyond}. This circuit is realized by a dc-SQUID, which acts as a flux-tunable inductor and, thus, changes the JPA resonance frequency $\omega_\text{SQZ}$ as a function of the applied local magnetic flux. The JPA resonator operates in the overcoupled limit and uses an input circulator which separates the input vacuum state from output squeezed states. In addition, the JPA has an independent microwave port, inductively coupled to the dc-SQUID, which is used to apply a strong parametric pump signal at the frequency $2\omega_\text{SQZ}$. More details about the design, fabrication and performance of the JPA can be found in Ref.\,\cite{honasoge2025}.

The propagating nonclassical squeezed signal is directed via a microwave cryogenic circulator to the spin-resonator hybrid device which resembles the quantum memory element in our setup. After interacting with the spin-resonator system, the reflected signal is amplified by a broadband low-noise cryogenic HEMT amplifier. The amplified output signal is processed further at room temperature by using a heterodyne detection setup with a high-frequency digitizer~\cite{menzel2010dual}. We calibrate gain and noise of the amplification chain, including the JPA properties, by making use of the Planck spectroscopy~\cite{mariantoni2010planck, Gandorfer2025}. This calibration procedure allows us to perform the Wigner tomography of our quantum states at various points of the cryogenic setup based on the reference state reconstruction approach~\cite{fedorov2018finite}.

The hybrid spin-resonator system is composed of a reflection type planar niobium microwave resonator and the electron spin ensemble, provided by an isotopically enriched $^{28}$Si crystal with a reduced nuclear spin population that is doped with phosphorus donors at a concentration of [P]~=~$1\text{x}10^{17}$~\si[per-mode=reciprocal]{\per\cubic\centi\meter}~\cite{weichselbaumer2020echo}. The $\SI{18}{\micro\meter}$ thick $^{28}\mathrm{Si}$ phosphorus-doped crystal is mounted in a flip-chip configuration onto the lumped element niobium resonator and held in place using a second silicon chip with a natural isotope composition acting as a cover. While phosphorus donors in silicon exhibit two electron spin resonance transitions due to the Fermi contact hyperfine interaction\,\cite{gordon1958microwave,feher1959electron}, we will selectively exploit only the $\ket{\downarrow\Downarrow}\rightarrow\ket{\uparrow\Downarrow}$ transition with frequency $\omega_\text{s}$. Here, we use the notation $\ket{\downarrow\Downarrow}$ to denote the energy levels, where the first (second) arrow corresponds to the electron (nuclear) spin. The energy levels are labeled according to their high-field product-state configuration (see also Appendix\,\ref{App: Spin properties}).

By controlling the static magnetic field $B_\text{0}$, provided by a superconducting solenoid magnet, $\omega_\text{s}$ can be tuned into resonance with the frequency of the resonator $\omega_\text{r}$. Notably, also the microwave resonator shows a weak frequency dependence with the applied in-plane magnetic field, which we attribute to the field dependent kinetic inductance of niobium. The JPA is spatially separated from the spin-resonator hybrid and placed outside the magnet. The additional superconducting magnetic shield guarantees close to quantum-limited performance. This setting allows to control the resonance frequency of the microwave resonator and the spin ensemble by tuning the external magnetic field $B_\text{0}$ independently of the JPA frequency. 

Ultimately, we can distinguish between three different scenarios: (i) a completely off-resonant regime, where the frequencies of the JPA, microwave resonator, and spin ensemble are all different, $\omega_{\text{SQZ}}\neq\omega_{\text{r}}\neq\omega_{\text{s}}$; (ii) a partially resonant regime, where the squeezed signal from the JPA (SQZ) is resonant with the microwave resonator, but the spin ensemble remains detuned, $\omega_{\text{SQZ}}=\omega_{\text{r}}\neq\omega_{\text{s}}$; (iii) a fully resonant regime, where all characteristic carrier frequencies of the aforementioned systems coincide, $\omega_{\text{SQZ}}=\omega_{\text{r}}=\omega_{\text{s}}$. These three distinct configurations allow us to unambiguously characterize all parts of our cryogenic setup and calibrate all important quantities independently, such as the squeezing level of the incoming microwave signals, the microwave response of the resonator, and the spectral characteristics of the spin-resonator hybrid system. \\

%-------------------------------------------------------------------
%\section{Theory}
%\label{sec: theory}
\textbf{Theoretical model.} Here, we briefly describe our approach to model the spin-resonator hybrid system consisting of a driven microwave resonator coupled resonantly to an electron spin ensemble within a steady-state, quantum-mechanical framework. It enables the analysis of the mean field amplitudes as well as the variances of microwave quantum signals. We define the Hamiltonian of the spin-resonator hybrid as
\begin{align}
\label{eq: Hamiltonian}
\begin{split}
      \hat{\mathcal{H}}/\hbar &= \omega_{\text{r}} \hat{a}^{\dagger}\hat{a} 
      + \omega_{\text{s}}\hat{s}^{\dagger}\hat{s}
      + g_\text{eff} \big( \hat{s}^{\dagger}\hat{a} +\hat{s}\hat{a}^{\dagger} \big)\\
      &+ i\sqrt{2\gamma_{\text{s}}} \big(\hat{s}^{\dagger}\hat{f}_{\text{s}} 
      + \text{H.c.} \big) 
      + i\big( \sqrt{2\kappa_{\text{int}}}\hat{b}_{\text{l}}\hat{a}^{\dagger}+ \text{H.c.}\big) \\
      &+ i\big( \sqrt{2\kappa_{\text{ext}}}\hat{b}_{\text{in}} \hat{a}^{\dagger}+ \text{H.c.}\big) 
      \text{,}
\end{split}
\end{align}
where we assume weak excitation signals and also neglect the inhomogeneous spin broadening~\cite{bienfait2017magnetic, wesenberg2009quantum, chiorescu2010magnetic, diniz2011strongly}. Here, $\hat{a}$ ($\hat{s}$) and $\hat{a}^{\dagger}$ ($\hat{s}^{\dagger}$) are the bosonic annihilation and creation operators for excitations inside the resonator (spin ensemble). The collective coupling between the spin ensemble and the resonator is defined by $g_{\text{eff}}=\sqrt{N}g_0$, with the number of spins $N$ and the single spin coupling rate $g_0$. The eigenfrequency of the resonator (spin ensemble) is $\omega_{\text{r}}$ ($\omega_{\text{s}}$). The spin dephasing is modeled using a collective coupling with rate $\gamma_\text{s}$ to a noise bath $\hat{f}_{\text{s}}$. Similarly, we account for internal resonator losses with an associated loss mode $\hat{b}_{\text{l}}$ and a loss rate $\kappa_{\text{int}}$. The coupling between the intra-cavity mode of the resonator and the external microwave circuit is described via the coupling of the input mode $\hat{b}_{\text{in}}$ with rate $\kappa_{\text{ext}}$. We assume all coupling rates as half width at half maximum linewidth. 

Using the Heisenberg equation of motion, we derive the following equations for the Fourier amplitudes of the resonator and spin operators
\begin{align}
\label{eq: eom resonator field}
\begin{split}
    -i\omega\hat{a}(\omega) &= -\big(i\omega_\text{r}+\kappa\big)\hat{a}(\omega) 
    + \sqrt{2\kappa_{\text{int}}}\hat{b}_{\text{l}}(\omega) \\
    &+ \sqrt{2\kappa_{\text{ext}}}\hat{b}_{\text{in}}(\omega)
    - i g_{\text{eff}}\hat{s}(\omega)\text{,}
\end{split}
\end{align}
\begin{align}
\label{eq: eom spin excitation}
    -i\omega\hat{s}(\omega) = -\big(i\omega_\text{s}+\gamma_\text{s}\big) \hat{s}(\omega)
     - i g_{\text{eff}}\hat{a}(\omega) + \sqrt{2\gamma_\text{s}}\hat{f}_\text{s}(\omega)\text{,}
\end{align}
with $\kappa=\kappa_{\text{int}}+\kappa_{\text{ext}}$. Following Ref.\,\cite{bienfait2017magnetic}, we obtain the relation between the input and output signal variances as
\begin{align}
\label{eq: input output variance}
    \sigma^2_{\hat{b}_\text{out}}(\omega) =  |r(\omega)|^2\sigma^2_{\hat{b}_\text{in}} + |l(\omega)|^2\sigma^2_{\hat{b}_\text{l}} + |t(\omega)|^2\sigma^2_{\hat{f}_\text{s}} \text{.}
\end{align}
Here, the variance of an operator $\hat{O}$ is defined as $\sigma_{\hat{O}}^2=\braket{\hat{q}^2_{\hat{O}}}$, with $\hat{q}_{\hat{O}}=(\hat{O}+\hat{O}^{\dagger})/2$. The complex coefficients $r$, $l$, and $t$ describe terms of the Langevin noise operator associated with the reflection from the cavity, the resonator internal losses, and losses of the spin ensemble, respectively. These coefficients are~\cite{bienfait2017magnetic}
\begin{align}
\begin{split}
     r(\omega) &= \text{Re}\bigg(1-\frac{2\kappa_\text{ext}}{X(\omega)}\bigg) \text{,}\\
    l(\omega)&= \text{Re}\bigg(\frac{\sqrt{4\kappa_\text{ext}\kappa_\text{int}}}{X(\omega)}\bigg) \text{,} \\
    t(\omega)&= \text{Re}\bigg(\frac{Y(\omega)}{X(\omega)} \bigg) \text{,}   
\end{split}
\end{align}
with the following definitions
\begin{align}
    \begin{split}
        X(\omega)&= -i\omega +\kappa + \frac{g_{\text{eff}}^{2}}{(\gamma_{\text{s}}+i(\Delta_\text{sr}-\omega))} \text{,}\\
        Y(\omega)&= \frac{\sqrt{4\gamma_{\text{s}}\kappa_\text{ext}}g_{\text{eff}}}{\gamma_{\text{s}}+i(\Delta_\text{sr}-\omega)} \text{.}
    \end{split}
\end{align}
Here, we define the detuning of the spin and resonator mode as $\Delta_\text{sr}=\omega_{\text{s}}-\omega_{\text{r}}$ and the detuning of the resonator mode and the incident microwave signal as $\Delta_\text{r}=\omega-\omega_{\text{r}}$. The condition $|l(\omega)|^2+|r(\omega)|^2+|t(\omega)|^2=1$ ensures that the output mode fulfills the bosonic commutation relation. We note that Eq.\,\ref{eq: input output variance} with the defined coefficients $r, l$ and $t$ is only valid at zero or large spin-resonator detunings\,\cite{thesisFesquet}. In addition, we compare the expectation value of the input and output fields within the input-output formalism and derive the complex scattering parameter~\cite{gardiner1985input, steck2007quantum, sandner2012strong}
\begin{align}
    \label{eq: scattering parameter}
    S_{11}(\omega) = \frac{\braket{\hat{b}_\text{out}}}{\braket{\hat{b}_\text{in}}} 
    = 1+\frac{2\kappa_{\text{ext}}}{i(\omega_{\text{r}}-\omega)-\kappa+\frac{g^2_{\text{eff}}}{i(\omega_{\text{s}}-\omega)-\gamma_{\text{s}}}} \text{.}
\end{align}

We analyze the output field variance in order to determine the squeezing level $S$ of the corresponding quantum state. To this end, we describe the output microwave signal in terms of its electromagnetic field quadrature operators, $\hat{q}$ and $\hat{p}$, with $\hat{b}_{\text{out}} = \hat{q}+i\hat{p}$, fulfilling the commutator relation $[\hat{q}, i\hat{p}]=1/2$. As derived in Ref.\,\cite{fedorov2018finite}, we can completely define the squeezed and anti-squeezed quadrature variances within this nomenclature. Consequently, the squeezing level can be expressed in decibels as
\begin{align}
    S = -10\,\text{log}_{10}(4\sigma_{\text{sq}}) \text{,}
\end{align}
where $\sigma_{\text{sq}}$ is the variance of the squeezed quadrature and positive values indicate squeezing below the vacuum level~\cite{honasoge2025}. \\

%----------------------------------------------------------------
%\section{Characterization of spin-resonator system}
\textbf{Characterization of the spin-resonator hybrid.} Next, we perform the continuous wave microwave spectroscopy of the resonator coupled to the electron spin resonance (ESR) transition. In detail, we measure the complex scattering parameter $S_{11}$ as a function of frequency using an incident microwave coherent signal with a power of $P_\text{mw} =-139$~dBm, corresponding to an average resonator occupation of $0.5$ photons at $\omega_{\text{r}}$~\cite{aspelmeyer2014cavity}. By controlling the applied static magnetic field $B_{0}$, we can tune the frequency of the ESR transition to match the resonator resonance frequency, $\omega_{\text{s}} = \omega_{\text{r}}$. For this condition, $S_{11}$ also reflects the conversion of coherent microwave photons to spin excitations.

\begin{figure}[t!]
    \centering
    \includegraphics[width=1\linewidth]{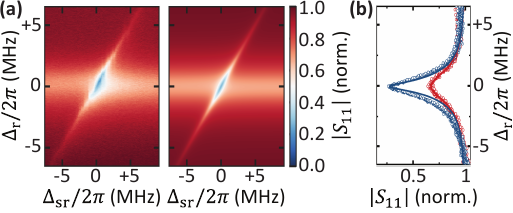}
    \caption{\textbf{Characterization of the spin-resonator hybrid.} (a) Measured (left) and modeled (right) microwave reflection amplitude $|S_{11}|$ as a function of the coherent probe-resonator detuning $\Delta_\text{r}$ and spin-resonator detuning $\Delta_\text{sr}$. The spectrum is modeled based on Eq.\,\ref{eq: scattering parameter}. (b) Comparison of the measured (dots) and modeled (lines) data for the resonant ($\Delta_{\text{sr}}/2\pi=0$) and off-resonant ($\Delta_{\text{sr}}/2\pi=\SI{9.2}{\mega\hertz}$) conditions, presented in blue and red, respectively.}
    \label{fig2: spin-resonator}
\end{figure}

\begin{figure*}[!t]
    \centering
    \includegraphics[width=1\textwidth]{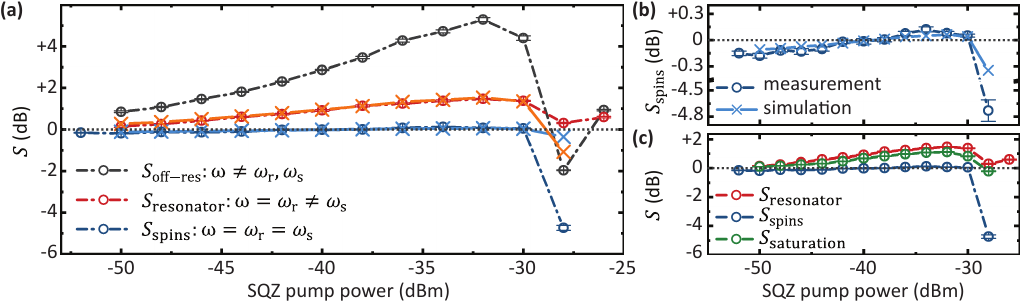}
    \caption{\textbf{Squeezing levels of propagating microwave signals after interaction with the spin-resonator hybrid.}(a) The measured squeezing level $S$ as a function of the SQZ pump power in the off-resonant case (black circles), in resonance with the resonator but off-resonant with the spins (red circles), and in resonance with the spin-resonator hybrid (blue circles). The lines are guides to the eyes. The crosses represent our model predictions based on Eq.\,\ref{eq: input output variance}, using the off-resonant squeezing states as reference input signals combined with independently determined coupling parameters $g_\text{eff}$, $\kappa_\text{ext}$, $\kappa_\text{int}$, and $\gamma_\text{s}$. (b) The measured and modeled $S$ in resonance with the spins indicate a squeezing below the vacuum level ($S>0)$. (c) Spin saturation measurements. A strong resonant microwave pulse saturates the ESR transition before performing a continuous wave squeezing measurement. This results in an increase of the squeezing levels approaching the reference experiment.}
    \label{fig3: cw squeezing}
\end{figure*}

Figure\,\ref{fig2: spin-resonator}(a) shows the measured microwave reflection amplitude $|S_{11}|$ close to $\omega_{\text{r}} \simeq \omega_{\text{s}}$ as a function of the detunings $\Delta_{\text{sr}}=(\omega_{\text{s}}-\omega_{\text{r}})\propto B_0-B_{\text{res}}$ and $\Delta_\text{r}=(\omega-\omega_{\text{r}})$. In the two-dimensional plot we identify two distinct features: an absorption line at $\Delta_{\text{r}}\approx 0$, which is independent of $\Delta_{\text{sr}}$, i.e. independent of the applied magnetic field, and is associated with the fixed-frequency microwave resonator; we also observe an ESR absorption signature, which varies linearly with the applied magnetic field, i.e. $\Delta_{\text{sr}}$. We can determine the characteristic parameters of both sub-systems by analyzing these spectral features. For $\Delta_{\text{sr}}/2\pi=\SI{9.2}{\mega\hertz}$, far detuned from the ESR line, we extract the resonance frequency $\omega_{\text{r}}/2\pi = \SI{5.645}{\giga\hertz}$ as well as the resonator linewidth $\kappa/2\pi=\SI{1.78}{\mega\hertz}$ and external coupling constant $\kappa_{\text{ext}}/2\pi=\SI{1.49}{\mega\hertz}$, using the circle fit method (see Appendix\,\ref{App: Data processing})~\cite{probst2015}. We next fit the total measured spectrum using Eq.\,\ref{eq: scattering parameter}, treating the aforementioned resonator parameters as known input parameters. In this way, we determine the intrinsic spin relaxation rate to $\gamma_{\text{s}}/2\pi=\SI{382}{\kilo\hertz}$ and the effective coupling rate to $g_{\text{eff}}=\SI{460}{\kilo\hertz}$. The experimentally determined cooperativity is $C = g_{\text{eff}}^2 / (\gamma_{\text{s}} \kappa) = 0.3$.

We compare the continuous wave ESR (cwESR) spectroscopy data with the response spectrum predicted by the input-output model of Eq.\,\ref{eq: scattering parameter} which we present in Fig.\,\ref{fig2: spin-resonator}(a). In addition, Fig.\,\ref{fig2: spin-resonator}(b) presents selected cross sections for the resonant and off-resonant cases of the spectra, demonstrating a quantitative agreement between the model prediction and the experimental data. We note a slight asymmetry in the experimentally measured absorption dips which is not described by our model and which is not corrected by the subtraction of the complex microwave background induced from imperfections of the experimental setup (see also Appendix\,\ref{App: Data processing}).\\

%-------------------------------------------------------
%\section{Continuous squeezing measurements}
\textbf{Probing the spin-resonator hybrid system with squeezed light.} Next, we experimentally investigate the reflection of propagating microwave squeezed states from the spin-resonator hybrid. To this end, we generate squeezed vacuum states using the squeezer JPA (SQZ), guide these nonclassical microwaves towards the input circulator of the spin-resonator system, and analyze the reflected signals using the reference state reconstruction protocol~\cite{fedorov2018finite}. We measure amplitudes of amplified microwave output signals using an field programmable gate array (FPGA) module with a detection bandwidth of $\SI{400}{\kilo\hertz}$. We use the FPGA internal logic to perform averaging and calculate the moments of the microwave signals up to the fourth order. This approach allows us to derive the Wigner function of the involved quantum states in the Gaussian approximation and estimate the squeezing level $S$ as a function of the applied pump power~\cite{fedorov2018finite, braunstein2005quantum}.

Figure\,\ref{fig3: cw squeezing}(a) shows the measured squeezing levels for the three different coupling scenarios: (i) $\omega_{\text{SQZ}}\neq\omega_{\text{r}}\neq\omega_{\text{s}}$, (ii) $\omega_{\text{SQZ}}=\omega_{\text{r}}\neq\omega_{\text{s}}$, and (iii) $\omega_{\text{SQZ}}=\omega_{\text{r}}=\omega_{\text{s}}$. Note, that we experimentally determine the squeezing level at the input of the HEMT amplifier (see Appendix\,\ref{App: Data processing} for details). All measurements are performed for a fixed frequency $\omega_\mathrm{SQZ}/2 \pi=\SI{5.645}{\giga\hertz}$. At $B_0 = 0$ the weak field dependence of the resonator results in a detuning of $(\omega_\mathrm{r}-\omega_\mathrm{SQZ})/2\pi \simeq \SI{10}{\mega\hertz}$~\cite{muller2022magnetic}, as compared to the spin ensemble working point at $\Delta_{\text{sr}}=0$. This results in a full reflection of the squeezed microwave state from the resonator input and allows us to obtain a reference baseline for the squeezing level as a function of the pump power. In this configuration, we observe a steady increase of $S$ with increasing pump power up to a maximum squeezing level of $S_\text{off-res, max}=(+5.29 \pm 0.09)$\,dB. Here, positive squeezing values indicate squeezing below the vacuum level. Above a pump power of $-30$\,dBm, the squeezing level abruptly decreases and, eventually, crosses the threshold value of $S = 0$\,dB. This could originate from the reconstruction model considering moments up to fourth order or from the disappearance of nonclassical correlations in the signal.
In the next step, we increase the static magnetic field, such that $\Delta_{\text{sr}}/2\pi = \SI{+9.2}{\mega\hertz}$. Then, the frequency of the SQZ and the resonator coincide while the spin mode remains detuned. In this configuration, the pump power dependence of the $S$ is qualitatively similar to the off-resonant reference measurement but shows a reduction in the squeezing level. The measured maximum squeezing level is $S_\text{r, max} = (+1.47 \pm 0.03)$\,dB. This reduction originates from the intrinsic resonator losses, as determined in the cwESR spectroscopy (see Fig.\,\ref{fig2: spin-resonator}(c)). These losses lead to an effective addition of thermal fluctuations from the coupled environmental bath, and consequently lead to an eventual reduction of the squeezing level.
Finally, we tune all our subsystems in resonance, $\omega_{\text{SQZ}} = \omega_{\text{r}} = \omega_{\text{s}}$, and measure the resulting squeezing level. We observe a further reduction of the power-dependent squeezing level as compared to the previous configurations. Here, the measured maximum squeezing level is $S_\text{spins, max} = (+0.12 \pm 0.03)$\,dB. We attribute this additional reduction of the squeezing level to the increased total absorption losses associated with the ESR transition.

For a quantitative analysis, we compare the presented experimental results with the prediction of the input-output model, as given by Eq.\,\ref{eq: input output variance}. Within this model, the variances in the measured output signal are determined by the variances of the reflected incoming squeezed microwave signals and thermal state contributions from the resonator and spin ensemble modes, which are assumed to be weak thermal baths. As an input signal variance $\sigma_{\hat{b}_\text{in}}$, we use the value obtained from the reference measurement, where $\omega_{\text{SQZ}} \neq \omega_{\text{r}} \neq \omega_{\text{s}}$. Using the coupling rates obtained from the earlier cwESR measurements, and describing the spin and resonator losses as arising from a thermal bath with the same temperature $T = \SI{70}{\milli\kelvin}$, we can use Eq.\,\ref{eq: input output variance} to estimate the expected squeezing level from the measured output variance. The calculated values are presented as crosses in Fig.\,\ref{fig3: cw squeezing}(a), showing an excellent quantitative agreement with the measured squeezing levels. Furthermore, the parameter $t$ in Eq.\,\ref{eq: input output variance} can be interpreted as a transfer efficiency between microwave signals and spin excitations, as it describes how strong thermal fluctuations of the spin systems couple to the microwave signal. Based on the fit of our model to the experimental data, we find $|t(0)|^2 \simeq 61\,\%$. 

\begin{figure}[!t]
    \centering
    \includegraphics[width=1\linewidth]{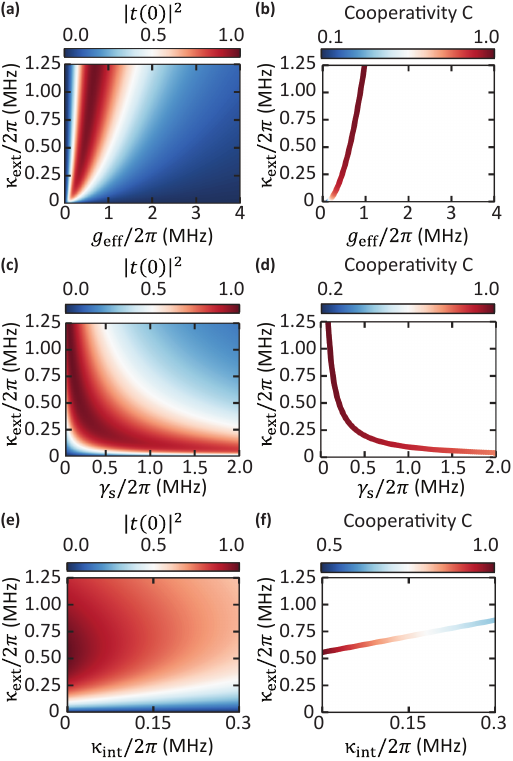}
    \caption{\textbf{Numerical calculation of the quantum transfer efficiency $|t(0)|^2$.} The transfer efficiency is computed according to Eq.\,\ref{eq: input output variance} as a function of different coupling strengths for the resonance case, i.e. $\omega_{\text{SQZ}} = \omega_{\text{r}} = \omega_{\text{s}}$. (a) Transfer efficiency as a function of the external coupling rate $\kappa_\text{ext}$ and effective spin-resonator coupling $g_\text{eff}$ for $\gamma_\text{s}/2\pi=\SI{382}{\kilo\hertz}$. (c) Transfer efficiency as a function of the external coupling rate $\kappa_\text{ext}$ and spin dephasing $\gamma_\text{s}$ for $g_\text{eff}/2\pi=\SI{460}{\kilo\hertz}$. The spin dephasing rate and the spin-resonator coupling are fixed to $\gamma_{\text{s}}=\SI{382}{\kilo\hertz}$ and $g_{\text{eff}}=\SI{460}{\kilo\hertz}$, respectively. (d) Transfer efficiency as a function of the external and internal coupling rates, $\kappa_\text{ext}$ and $\kappa_\text{int}$, respectively. Optimal state transfer and maximum transfer efficiency is achieved in the limit of unit cooperativity, $C = 1$. Panels (b), (d), and (f) illustrate the calculated cooperativity for the maximum coupling efficiency.}
    \label{fig4: simulation of efficiency}
\end{figure}

Interestingly, besides detuning the subsystems there is another method to decouple the spin ensemble from the microwave resonator. Conceptually, the application of a strong resonant microwave pulse can saturate the spin transition and decouple the resonator from the spin ensemble. In order to experimentally investigate this regime, we operate in the resonant configuration between all three subsystems and additionally apply a resonant, strong, rectangular-shaped, saturation pulse at the carrier frequency of $\omega/2\pi = \SI{5.645}{\giga\hertz}$ for the duration of $\SI{20}{\milli\second}$ with the power of $-120$\,dBm before performing the continuous wave measurements with the squeezed states. Within our squeezing measurements, we choose a pulse repetition time of $\SI{2}{\second}$. The measured squeezing level at the output of the spin-resonator hybrid is shown in Fig.\,\ref{fig3: cw squeezing}(c) with the maximum squeezing level of $S_\text{spins, sat} = (+1.12 \pm 0.03)$\,dB. This value closely reaches the originally measured maximal squeezing level for the case of the SQZ in resonance with the resonator, without the saturation pulse. The clear increase in squeezing level thus confirms that the saturation pulse leads to a decoupling of the spin ensemble and the microwave resonator. Quantitatively, the squeezing levels do not reach the reference values (red line in Fig.\,\ref{fig3: cw squeezing}). We attribute this observation to two aspects: the nonlinear properties of the spin-resonator hybrid and the timescales of the saturation protocol, i.e. the finite $T_1$ time of the donor spin ensemble which counteracts the spin-resonator decoupling. These nonlinear properties require a more complex description going beyond the simple model provided by Eq.\,\ref{eq: Hamiltonian}. \\
%\section{Simulation of the transfer efficiency}
\textbf{Simulation of the quantum transfer efficiency.} In order to investigate optimal coupling parameters for the spin-resonator hybrid platform, we numerically compute the transfer efficiency using Eq.\,\ref{eq: input output variance}. We study experimentally reachable parameter regimes for realistic internal and external resonator coupling strengths, a spin dephasing rate, and an effective spin-resonator coupling. Figure\,\ref{fig4: simulation of efficiency} presents the results of those numerical calculations which provide a foundation for the realization of an optimal quantum state transfer for future experiments. Here, we investigate the resonant case, $\omega_{\text{SQZ}} = \omega_{\text{r}} = \omega_{\text{s}}$, and thus investigate the transfer efficiency $|t(0)|^2$ as a function of different coupling rates.

In the first case, we choose the spin dephasing rate to be fixed, $\gamma_{\text{s}}=\SI{382}{\kilo\hertz}$, and vary the external coupling rate of the resonator as well as the effective spin-resonator coupling. Panels (a) and (b) in Fig.\,\ref{fig4: simulation of efficiency} present the estimated effective transfer strength and cooperativity corresponding to a maximized transfer efficiency, respectively. We observe a parameter regime with an increasing coupling efficiency, approaching unity in the case of $\kappa_\text{ext}/2\pi = \SI{1.25}{\mega\hertz}$ and $g_\text{eff}/2\pi = \SI{976}{\kilo\hertz}$. For these parameters, the corresponding cooperativity is reaching $C_{\text{max}}=0.982$. Next, we fix the spin-resonator coupling, $g_{\text{eff}}=\SI{460}{\kilo\hertz}$, and vary the external coupling of the resonator as well as the spin dephasing rate. Panels (c) and (d) in Fig.\,\ref{fig4: simulation of efficiency} show the resulting coupling regime and corresponding cooperativity. The coupling efficiency reaches value of $0.991$ for $\kappa_\text{ext}/2\pi=\SI{1.25}{\mega\hertz}$ and $\gamma_\text{s}/2\pi=\SI{167}{\kilo\hertz}$, corresponding almost to unit cooperativity. In the third scenario, we keep the spin dephasing rate and effective spin-resonator coupling constant and equal to the determined experimental values, i.e. $\gamma_{\text{s}}=\SI{382}{\kilo\hertz}$ and $g_{\text{eff}}=\SI{460}{\kilo\hertz}$. Then, we vary the internal and external coupling rates of the resonator, which can experimentally be achieved by altering the resonator design. We find a maximum coupling efficiency for $\kappa_{\text{int}}/2\pi = \SI{0.25}{\kilo\hertz}$ and $\kappa_{\text{ext}}/2\pi=\SI{556}{\kilo\hertz}$ with the corresponding cooperativity of $C=0.999$. 

Our simulations of the coupling efficiency for various coupling parameters of our spin-resonator hybrid system are in agreement with results reported literature as can be found in Ref.\,\cite{afzelius2013proposal, julsgaard2013quantum, greggio2025optimal}. We want to highlight, that our spin-resonator hybrid system is engineered with a cooperativity of $C=0.3$ for which we expect a high coupling efficiency. In order to further increase the coupling efficiency of quantum states to our spin ensemble, we can tune the coupling parameters of our microwave resonator. One way can be to improve the internal quality factor, e.g. by using magnetic field robust superconducting materials like NbTiN. Furthermore, we can adapt the design of the microwave circuit to tune the coupling of incoming quantum states to the microwave resonator. \\

\textbf{Conclusion.} In summary, we have studied a spin-resonator hybrid system interacting with squeezed microwave signals at various conditions. We have observed a strong transfer of the squeezed microwaves to the spin ensemble corresponding to a transfer efficiency of $|t|^2 \simeq 61\,\%$. We have successfully modeled our experimental findings by an input-output quantum formalism, which also allows us to extrapolate the resonator properties required to reach the cooperativity of unity, as demanded by the optimal state transfer with spin ensembles~\cite{bernad2025analytical, greggio2025optimal}. Our results provide new insight in the fundamental understanding of  solid-state spin systems driven with nonclassical signals. In particular, we expect such systems to become useful for quantum memory applications in microwave quantum communication~\cite{Fesquet2024}. This is highly relevant for quantum microwave communication and quantum information processing with superconducting circuits. \\

\begin{acknowledgments}
We acknowledge support by the German Research Foundation via Germany's Excellence Strategy (EXC-2111-390814868), the German Federal Ministry of Education and Research via the project QuaMToMe (Grant No. 16KISQ036), and by the Deutsche Forschungsgemeinschaft via the Transregio ConQuMat TRR 360 (Project-ID 492547816). This research is part of the Munich Quantum Valley (lighthouse project IQ-Sense), which is supported by the Bavarian state government with funds from the Hightech Agenda Bayern Plus.
\end{acknowledgments}

\section*{Data availability}
The data displayed is openly available: \url{https://doi.org/10.5281/zenodo.17912272}.
\newpage
\appendix
\section{Spin properties}
\label{App: Spin properties}
%\subsection{Phosphorus donors in silicon}
\textbf{Phosphorus donors in silicon.} A single phosphorus donor in silicon provides an unbound electron with electron spin $S = 1/2$ and a nuclear spin from the $^{31}$P nuclei of $I = 1/2$. Consequently, a corresponding spin Hamiltonian of the spin ensemble in an external magnetic field can be expressed as \cite{schweiger2001principles}:
\begin{align}
    \hat{\mathcal{H}}_{\text{s}} =  g_\text{e} \mu_\text{B} \hat{B_{\text{0}}}\hat{S} + g_\text{n} \mu_\text{n} \hat{B_{\text{0}}}\hat{I}  + A \hat{S} \hat{I} \text{,}
\end{align}
where the first two terms are the Zeeman interaction of the electron and nuclear spin systems, respectively. The last term represents the Fermi-contact hyperfine interaction with the hyperfine constant $A/h=\SI{117.53}{\mega\hertz}$. The g-factors of the electron and nuclear spins are $g_{\text{e}}=1.9985$ and $g_{\text{n}}=2.2632$, respectively. Furthermore, $\mu_{\text{B}}$ and $\mu_{\text{n}}$ are the Bohr and nuclear magnetons, respectively.

In the limit of high magnetic fields ($g_\text{e} \mu_\text{B} {B_{\text{0}}} \gg A$), we expect two spin resonant transitions. In addition to these desired ESR transitions, signatures of dangling bond defects located at the Si/SiO$_2$ interface are additionally visible in the spectrum (see e.g. \cite{weichselbaumer2019, zollitschapl}). The electron spin sub-ensemble linked to the nuclear spin-down configuration is known to avoid the overlap of multiple ESR transitions and is thus investigated here. \\

%\subsection{Pulsed ESR spectroscopy}
\textbf{Pulsed ESR spectroscopy.} We perform a pulsed ESR spectroscopy based on Hahn-echo sequences to determine the characteristic coherence times of our spin ensemble \cite{hahn1950spin, tyryshkin2003electron, schweiger2001principles}. We rely on the Zeeman shift of the spin mode in resonance with the coupled cavity mode. Then, we generate in-phase and out-of-phase signals of Gaussian-shaped pulses using an arbitrary waveform generator and modulate carrier signals at the pulses to the resonance frequency of our spin-resonator hybrid using a vector signal generator. The detection of the resulting spin echo is performed by a heterodyne down-conversion setup. We analyze the area of the spin echo $A_{\text{echo}}$ by fitting a Gaussian function to the data and integrating the area within a time window around the maximum amplitude of the echo given by 3 times the standard deviation.

\begin{figure}[!t]
    \centering
    \includegraphics[width=1\linewidth]{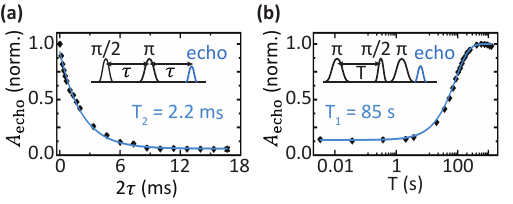}
    \caption{\textbf{Coherence times of the Si:P ensemble.} Experimentally determined coherence times $T_2$ (a) and $T_1$ (b) based on conventional Hahn echo and inversion recovery pulse sequences, respectively.}
    \label{fig: appendix coherence times}
\end{figure}

Furthermore, we optimize the Gaussian input pulses by using detected echo areas. We choose the width of the $\pi/2$-pulse and $\pi$-pulse to be equal, but the pulse amplitudes to be different. We find a maximum spin echo area by choosing a pulse width of $\SI{2}{\micro\second}$ and amplitude ratio of $0.625/1$ between the $\pi/2$-pulse and $\pi$-pulse. Using those optimized pulse parameters, we vary the delay time $\tau$ between the two incoming pulses to determine the characteristic coherence time $T_{2}$ within a conventional Hahn-echo pulse sequence, $\pi/2$-$\tau$-$\pi$-$\tau$-echo. Figure\,\ref{fig: appendix coherence times}(a) shows the resulting echo area, which is exponentially decreasing for increasing delay times. We fit an exponential decay function to the experimental data, resulting in $T_{2}=\SI{2.16\pm0.11}{\milli\second}$. Similarly, we perform an inversion recovery pulse sequence, $\pi$-$T$-$\pi/2$-$\tau$-$\pi$-$\tau$-echo, to determine the characteristic coherence time $T_{1}=\SI{85.49\pm0.99}{\second}$.

%\subsection{Experimental setup}
\section{Experimental setup}
\begin{figure}[!t]
    \centering
    \includegraphics[width=\linewidth]{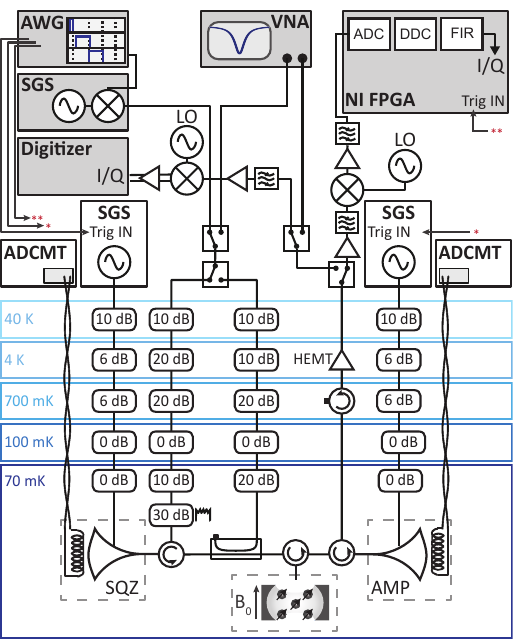}
    \caption{\textbf{Experimental setup.} Cryogenic and room-temperature setup for continuous wave and pulsed measurements with coherent microwave signals as well as the generation and detection of squeezed microwave signals. Further details are given in the text.}
    \label{fig: appendix setup}
\end{figure}

\textbf{Cryogenic wiring.} The experimental setup is shown in Fig.\,\ref{fig: appendix setup}. Within the cryogenic microwave circuit, we are equipped with a total of four input lines with different attenuation to suppress room temperature noise. Two of the input lines serve as pump lines for the JPAs. To probe our measurement devices, we have an input line with a total of $90$~dB attenuation that is coupled to the SQZ via a cryogenic circulator (Low Noise Factory LNF-CIC4\_8A). We can bypass the SQZ and probe the spin ensemble via an additional input line with $60$~dB attenuation, a cryogenic directional coupler (Sirius Microwave CPL2000-18000-30-C), and an additional cryogenic circulator (Low Noise Factory LNF-CIC4\_8A). The output signal is guided trough a cryogenic amplification chain via another circulator (Low Noise Factory LNF-CIC4\_12A) to a second JPA  (AMP) and ultimately via a circulator (Quinstar CTH0408KCS) to the cryogenic HEMT amplifier (Low Noise Factory LNF-LNC4\_8C) at the 4K stage. Note, that for the experiments presented in this work we did not use the second JPA (AMP) for the amplification of the output signals. As described before, the spin-resonator device is placed in the center field of a solenoid magnet, while both JPAs are placed outside the magnet. \\

\textbf{Room temperature setup.} At room temperature, we can use a total of three different measurement techniques. For continuous wave measurements, we probe our devices with a vector network analyzer (Rohde \& Schwarz ZVA8). For pulsed ESR measurements, we use an arbitrary waveform generator (Zurich Instruments HDAWG8) to generate in-phase and out-of-phase signals with Gaussian-shaped pulses. The pulses are up-converted to the resonance frequency using a vector signal generator (Rhode \& Schwarz SGS100A). For the detection of spin echoes, the output signal is filtered, amplified, and down-converted using an IQ mixer (Marki IQ-0307L). Then, we use an additional variable amplifier (FEMTO DHPVA-200) and digitize the pulsed signal with an analog-to-digital converter (Spectrum M4i.4451-x8). For the generation of squeezed microwave signals, we use a DC voltage source (ADCMT 6241A) to apply a flux bias setting the resonance frequency of the JPA and a microwave signal generator (Rohde \& Schwarz SGS100A) for the generation of the pump tone. The detection of the output signal is performed in a heterodyne detection setup, including a variety of bandpass filters, room-temperature amplifiers and the FPGA module (National Instruments FPGA 7972).

%\subsection{Background correction cwESR}
\section{Data processing and calibration techniques}
\label{App: Data processing}
\textbf{Background correction of the cwESR spectrum.} When measuring cwESR spectra, the measured scattering parameter is subject to the complex microwave background based on imperfections in the experimental setup, such as impedance mismatches and signal delays due to finite cable lengths. To correct for those effects, we apply the circle-fit routine and fit the resonator response detuned from the spin mode, $\Delta_{\text{sr}}/2\pi = \SI{9.2}{\mega\hertz}$, using the following equation \cite{probst2015}:
\begin{align}
    |S_{\text{11, exp}}| = \Bigg|A_{\text{MW}} \left(1 - \frac{2Q_{\text{l}} / Q_{\text{ext}}}{1 + 2i Q_{\text{l}} (\omega / \omega_{\text{r}} - 1)} \right)\Bigg| \text{,}
    \label{eq: S11 resonator}
\end{align}
where the complex amplitude $A_{\text{mw}}=a_{\text{mw}}\text{e}^{i\left(\alpha-\omega\tau\right)}$ captures environmental influences, including amplitude variations, impedance-induced phase shifts, and microwave signal propagation delays. The loaded quality factor $Q_\text{l}$ is defined as the sum of reciprocal values of the internal and external quality factors ($Q_\text{int}$ and $Q_\text{ext}$), $Q_\text{l}^{-1}=Q_\text{int}^{-1}+\text{Re}(Q_\text{ext}^{-1})$. The internal quality factor accounts for intrinsic losses such as radiative losses, surface resistance, magnetic field-induced dissipation, and spin ensemble interactions. The complex external quality factor $Q_{\text{ext}} = |Q_{\text{ext}}|\text{e}^{-i\phi}$ quantifies energy dissipation into the external feedline, while the phase $\phi$ accounts for impedance mismatches and parasitic effects~\cite{probst2015}. Applying the circle-fit routine, we find $a_{\text{mw}}=0.021$, $\alpha=1.585$, $\tau=\SI{97.35}{\nano\second}$, $\omega_\text{r}/2\pi =\SI{5.645}{\giga\hertz}$, $\phi=-0.002$, $Q_\text{ext}=(1900\pm 11)$, and $Q_\text{l}=(1586\pm 15)$.

Assuming, that the environmental influences are independent on the applied magnetic field for the investigated field range, we correct our measured data as follows
\begin{align}
    |S_\text{11, corr}| = \Bigg|\frac{\text{Re}(S_\text{11, meas})+i~\text{Im}(S_\text{11, meas})}{a_{\text{mw}} \text{e}^{i(\alpha-\omega\tau)}}\Bigg| \text{.}
\end{align}
Furthermore, we use a normalization of the corrected data in order to compare the measured spectrum to the theoretical model. The resulting scattering parameter is defined as
\begin{align}
    |S_{11}| = |S_\text{11, corr}| / \text{max}(|S_\text{11, corr}|) \text{.}
\end{align}
%\textbf{Magnetic field calibration.} In addition to the correction of the measured scattering parameter, we have to correct our spectrum for magnetic off-set fields. Such additional field contributions may arise from the earth magnetic field or electromagnetic sources in our lab close to the cryostat. To calibrate the strength of the static magnetic field $B_0$ at the sample position within the experimental setup, we rely on the two ESR transitions of our Si:P sample. We find an offset magnetic field strength of $\Delta B=B_{\text{set}}-B_0=\SI{1.405}{\milli\tesla}$ for which we account in the data presentation. \\
%\subsection{Reference state reconstruction}
\textbf{Reference state reconstruction.}
As described in the main text, we digitize the microwave output signal at room temperature with a heterodyne detection setup using an FPGA module. With the recorded $I/Q$ sampled data points, we compute the quadrature moments up to the fourth order, $\braket{I^n Q^m}$ for $n + m \le 4 ((n,m)\in \mathbb{N} ^2)$, and consequently we compute the corresponding signal moments with respect to a specific reconstruction point in our setup. Thus, we can reconstruct the squeezing angle and squeezing level for the detected output signal~\cite{fedorov2018finite, eichler2011observation}. 
In order to map the detected voltage amplitude of the microwave output signal to a photon number, we rely on Planck spectroscopy~\cite{mariantoni2010planck, Gandorfer2025}. Here, we use a heatable input attenuator that is placed within our microwave circuit before the SQZ. Then, we measure the output voltage dependent on the temperature of the heatable attenuator. Based on a Planck's distribution fitting routine, we can determine a photon number conversion factor which relates the voltage measured at room temperature to a photon number at the desired reference point in the cryogenic setup. Within our data analysis, we choose the HEMT input as the reconstruction point. To this end, we carefully estimate the microwave losses between the heatable attenuator and the HEMT input to be $-2.8$~dB.

%\subsection{Characterization of the squeezer JPA}
\section{Characterization of the squeezer JPA}
\begin{figure}[!t]
    \centering
    \includegraphics[width=\linewidth]{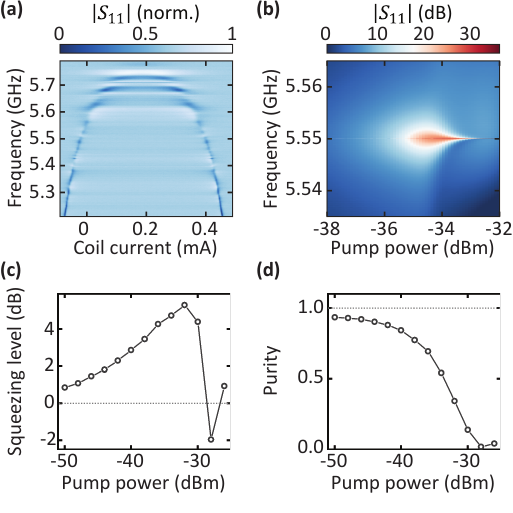}
    \caption{\textbf{Experimental characterization of the squeezer JPA.} (a) Measurement of the microwave reflection amplitude $|S_{11}|$ as a function of the probe frequency and dc current through the superconducting magnetic field coil. (b) Measurement of the dependence of the non-degenerate gain on the applied pump power for a fixed frequency of $\omega_{\text{pump}}/2\pi = \SI{11.1}{\giga\hertz}$. (c) Measured squeezing level as a function of the applied pump power for a fixed frequency of $\omega_{\text{SQZ}}/2\pi = \SI{5.645}{\giga\hertz}$ showing a maximum squeezing level of $(+5.29 \pm 0.09)$~dB. The gray-dashed line depicts the vacuum limit, above which the reconstructed microwave signals are referred to as squeezed. (d) The corresponding reconstructed purity dependent on the applied pump power. The gray-dashed line illustrates the upper limit at which the purity becomes unity.}
    \label{fig: appendix JPA measurements}
\end{figure}

\textbf{Flux-dependent resonance frequency and non-degenerate gain.} Here, we present characterization measurements of the squeezer JPA (SQZ). First of all, we record the dependence of the resonance frequency on the applied magnetic flux. The result is shown in Fig.\,\ref{fig: appendix JPA measurements}(a). In our experiment, we measure the complex microwave reflection amplitude $|S_{11}|$ as a function of the probe frequency and dc current through the superconducting magnetic field coil. We observe a maximum resonance frequency of $\omega_{\text{}}/2\pi = \SI{5.73}{\giga\hertz}$ and a tunability range of about \SI{500}{\mega\hertz}. Importantly, we identify two coil currents at which we can tune the SQZ in resonance with our spin-resonator hybrid, fulfilling $\omega_{\text{SQZ}} = \omega_{\text{r}} = \omega_{\text{s}}$. Furthermore, we measure the dependence of the non-degenerate gain on the applied pump power for a fixed frequency of $\omega_{\text{pump}}/2\pi = 2\,\omega_{\text{SQZ}}/2\pi = \SI{11.1}{\giga\hertz}$. The result is shown in Fig.\,\ref{fig: appendix JPA measurements}(b). We record a maximum non-degenerate gain of $G_{\text{nd}}=36$~dB for a fixed power of the coherent pump tone of $-32.3$~dBm.\\

\textbf{Generation of squeezed microwave signals.} Finally, we characterize the generation of squeezed microwave signals by recoding the dependence of the squeezing level on the applied pump power for a fixed frequency of $\omega_{\text{SQZ}}/2\pi = \SI{5.645}{\giga\hertz}$. The results are shown in Fig.\,\ref{fig: appendix JPA measurements}(c). We observe a maximum squeezing level of $(+5.29 \pm 0.09)$~dB. Correspondingly, we reconstruct the purity as a function of the applied pump power. The results are shown in and Fig.\,\ref{fig: appendix JPA measurements}(d). With increasing pump power, the purity steadily decreases due to pump-induced noise, higher-order nonlinearities and gain-dependent environmental noise\,\cite{boutin2017effect, renger2021beyond}.
We use this power-dependent squeezing measurement as a reference for the coupling of squeezed microwave signals to the spin-resonator hybrid system. \\

%----------------------------------------------------
%----------------------------------------------------
\newpage
%\bibliography{Bibliography}

\end{document}